\documentclass[conference]{IEEEtran}

\usepackage{color}
\usepackage{cite}
\usepackage{fixltx2e}
\usepackage[cmex10]{amsmath}
\usepackage{array}
\usepackage{wasysym}
\usepackage{dsfont}
\usepackage[latin1]{inputenc}
\usepackage[english]{babel}
\usepackage[T1]{fontenc}
\usepackage{mathtools}
\usepackage{amssymb}
\usepackage{enumerate}
\usepackage{bbm}
\usepackage{epsfig,syntonly}
\usepackage{verbatim,times}
\usepackage{epstopdf}
\usepackage{graphicx}
\usepackage{tikz}
\usepackage{pgfplots}
\usepackage{wrapfig}
\usepackage{graphicx} 
\usepackage{latexsym,fancyhdr,bm}
\usepackage{url}
\usepackage{bbm}
\usepackage{subcaption}

\usepackage{amsthm}

\newtheorem{theorem}{Theorem}

\newtheorem{lemma}{Lemma}
\newtheorem{prop}{Proposition}

\newcommand{\code}{{\mathcal C}}

\newcommand{\cw}{x}
\newcommand{\argmax}{\text{argmax}}

\newcommand{\erasure}{\omega}

\makeatletter
\let\l@ENGLISH\l@english
\makeatother

\title{Almost Optimal Scaling of Reed-Muller Codes on BEC and BSC Channels}
\author{\IEEEauthorblockN{Hamed Hassani\IEEEauthorrefmark{1}, Shrinivas Kudekar\IEEEauthorrefmark{2}, Or Ordentlich\IEEEauthorrefmark{3}, Yury Polyanskiy\IEEEauthorrefmark{4}  and R\"{u}diger Urbanke\IEEEauthorrefmark{5}}
\IEEEauthorblockA{\IEEEauthorrefmark{1}University of Pennsylvania, Email: \texttt{hassani@seas.upenn.edu}}
\IEEEauthorblockA{\IEEEauthorrefmark{2}Email: \texttt{kudekar@gmail.com}}
\IEEEauthorblockA{\IEEEauthorrefmark{3}Hebrew University of Jerusalem, Israel, Email: \texttt{or.ordentlich@mail.huji.ac.il}}
\IEEEauthorblockA{\IEEEauthorrefmark{4}MIT EECS, USA, Email: \texttt{yp@mit.edu}}
\IEEEauthorblockA{\IEEEauthorrefmark{5}School of Computer and Communication Sciences, EPFL, Switzerland, Email: \texttt{ruediger.urbanke@epfl.ch}}}

\begin{document}

\maketitle
\begin{abstract}
\noindent
Consider a binary linear code of length $N$, minimum distance
$d_{\text{min}}$, transmission over the binary erasure channel with
parameter $0 < \epsilon < 1$ or the binary symmetric channel with
parameter $0 < \epsilon < \frac12$, and block-MAP decoding.  It was
shown by Tillich and Zemor that in this case the error probability
of the block-MAP decoder transitions ``quickly'' from $\delta$ to
$1-\delta$ for any $\delta>0$ if the minimum distance is large. In
particular the width of the transition is of order
$O(1/\sqrt{d_{\text{min}}})$. We strengthen this result by showing
that under suitable conditions on the weight distribution of the
code, the transition width can be as small as
$\Theta(1/N^{\frac12-\kappa})$, for any $\kappa>0$, even if the
minimum distance of the code is not linear. This condition applies
e.g., to Reed-Mueller codes. Since $\Theta(1/N^{\frac12})$ is the
smallest transition possible for any code, we speak of ``almost''
optimal scaling.  We emphasize that the width of the transition
says nothing about the location of the transition. Therefore this
result has no bearing on whether a code is capacity-achieving or
not. As a second contribution, we present a new estimate on the
derivative of the EXIT function, the proof of which is based on the Blowing-Up Lemma.
 \end{abstract}


\section{Introduction}
Consider a binary linear code of length $N$ and minimum distance
$d_{\text{min}}$. Assume that we transmit over the binary erasure
channel (BEC) with parameter $\epsilon$, $0 < \epsilon < 1$, or the
binary symmetric channel (BSC) with parameter $\epsilon$, $0 <
\epsilon < \frac12$. Assume further that the receiver performs block
maximum-a posteriori (block-MAP) decoding.  It was shown by Tillich
and Zemor \cite{tillich2000discrete} that in this case the error
probability transitions ``quickly'' from $\delta$ to $1-\delta$ for
any $\delta>0$ if the minimum distance is large. In particular they
showed that the width of the transition is of order
$O(1/\sqrt{d_{\text{min}}})$. For codes whose minimum distance is
proportional to the blocklength, this gives a transition width of
$O(1/\sqrt{N}))$ and this is the best possible. But for codes whose
minimum distance is sublinear the width ``guaranteed'' by this
result is sub-optimal. E.g, if we consider Reed-Mueller (RM) codes
of fixed rate and increasing length, then their minimum distance
grows only like $\Theta(\sqrt{N})$.

In this paper, we strengthen this scaling result. We show that under suitable
conditions on the weight distribution of the code, the transition
width will be nearly optimal, i.e., it will be as small as
$\Theta(1/N^{\frac12-\kappa})$, for any $\kappa>0$.  The required
condition applies e.g., to RM codes, and hence we see that RM codes
have an almost optimal scaling of their block error probability
under block-MAP decoding.

It is important to note that the width of the transition has no
bearing on {\em where} this transition happens. This is analogous
to concentration results in probability (think of Azuma's inequality)
where one can prove that a random variable is concentrated around
its mean without determining the value of the mean.  Therefore this
result has no bearing on whether a code is capacity-achieving or
not. In particular, our result does not resolve the question whether
RM codes are capacity-achieving over any channel other than the
BEC, see \cite{kudekar2017reed,kudekar2016comparing}.

Moreover, even though RM codes are known to achieve capacity over
the BEC, our results do not imply that the gap to capacity of these
codes at a fixed error probability scales like $O(1/\sqrt{N})$. The
reason being that~\cite{kudekar2017reed} only shows that a sharp
transition occurs at $\epsilon^*>1-C-O(1/\log {N})$. To establish
the desired $O(1/\sqrt{N})$ gap-to-capacity result for RM over the BEC,
one would need to obtain tighter bounds on $\epsilon^*$. As a first
step in this direction, we develop a new tool for estimating the
derivative of the EXIT function. Roughly speaking, we show that,
for a transitive code, if for most pairs of erased locations $(i,j)$
for which $H(X_i,X_j|Y_{\sim i,j})=1$ bit, the conditional probability
of the event $X_i=X_j=1$ does not decrease with $N$, then the EXIT
function transitions sharply with a transition width of $O(1/\sqrt{N})$.
While we are currently unable to verify this condition for RM codes
analytically, numerical indications suggest that this might indeed
be the case. Our new estimate on the EXIT function derivative is based on the 
Blowing-Up Lemma.

\section{Preliminaries}
\noindent\textbf{Linear Codes.} Let $\code$ be a binary linear code
of length $N$, dimension $K$, and minimum distance $d_{\min}$.  We
let $A(w)$ denote the weight distribution function of $\code$, i.e.,
for any $w \leq N$ we have
\begin{equation}
A(w) := \left| \left\{ x \in \code: w_H(x) = w  \right\} \right|,
\end{equation}
where $w_H(x)$ denotes the Hamming weight of vector $x$. Let us also define the function $A(W, z)$ as follows:
 \begin{equation}
 A(W, z) \triangleq \sum_{w=1}^W A(w) z^w.
 \end{equation}
\noindent\textbf{Transmission Channel and Block-MAP Decoding}
We consider transmission over two types of channel families: the
binary erasure channel with parameter $\epsilon$ (BEC($\epsilon$))
and the binary symmetric channel with cross-over probability
$\epsilon$ (BSC($\epsilon$)).  Let $X$ be the codeword, chosen
uniformly at random from $\code$, and let $Y$ be the received word.
When transmission is over the BEC we have $Y \in \{0, 1, ?\}^N$,
and when it is over the BSC we have $Y \in \{0,1\}^N$. Let $\hat{x}(y)$
be the block-MAP decoding function
\begin{align}
\hat{x}(y) = \argmax_{x \in \code} p(y | x),
\end{align}
where ties are resolved in an arbitrary fashion.

We let $P_{\rm MAP}(\epsilon)$ denote the probability of error for
the block-MAP decoder, i.e., $P_{\rm MAP}(\epsilon)=P\{\hat{x}(Y)
\neq X\}$. Here, to simplify notation, we have used the same notation
(i.e., $P_{\rm MAP}(\epsilon)$) for transmission over both the
BEC($\epsilon$) and the BSC($\epsilon$), and in the sequel, the
choice of the transmission channel will be clear from the context.

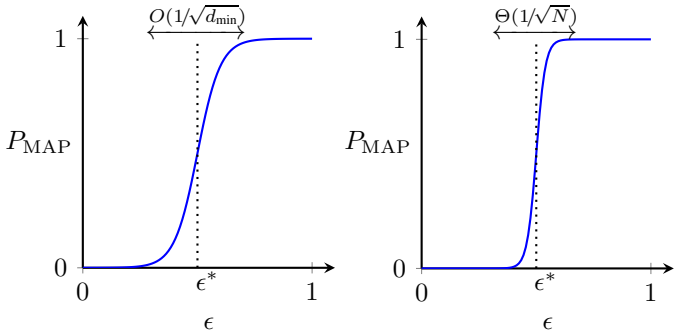
\begin{figure}[htb]
  \centering
  \begin{subfigure}{0.15\textwidth}
    \begin{center}

\begin{tikzpicture}
\hspace{-2cm}
 \begin{axis}[scale=1.2, thick,
width=1.1in,
height=1.1in,
scale only axis,
axis y line=left,
axis x line=bottom,
font=\normalsize,
xmin=0,
xmax=1.1,
xlabel={$\epsilon$},
ymin=0,
ymax=1.1,
ylabel={$P_{\rm MAP}\hspace{-1.3cm}$},
ylabel style={rotate=-90},
ytick={0,1},
xtick={0,1},
axis x line*=bottom,
axis y line*=left,
legend style={at={(.415,.1)},anchor=south west, legend cell align=left, font=\normalsize}
]
   
    \addplot [mark=none,samples=100,domain=0:1,color=blue,thick] {1/2*(1+tanh(10*(\x-1/2)))};

    \draw [dotted, thick] (axis cs:1/2,0)-- (axis cs:1/2,1);
\end{axis}
\node[label=below:{$\epsilon^*$}] at (1.65,0.2) {};
\node[label=above:{$\xleftrightarrow{\!\!O(1\!/\!{\sqrt{d_\text{min}}})\!\!}$}] at (1.5,2.8) {};

\end{tikzpicture}
    \end{center}
  \end{subfigure}%
 \,\,
  \begin{subfigure}{0.15\textwidth}
    \begin{center}

\begin{tikzpicture}
\hspace{-.5cm}
 \begin{axis}[scale=1.2, thick,
width=1.1in,
height=1.1in,
scale only axis,
axis y line=left,
axis x line=bottom,
font=\normalsize,
xmin=0,
xmax=1.1,
xlabel={$\epsilon$},
ymin=0,
ymax=1.1,
ylabel={$P_{\rm MAP}\hspace{-1.3cm}$},
ylabel style={rotate=-90},
ytick={0,1},
xtick={0,1},
axis x line*=bottom,
axis y line*=left,
legend style={at={(.415,.1)},anchor=south west, legend cell align=left, font=\normalsize}
]
   
    \addplot [mark=none,samples=100,domain=0:1,color=blue,thick] {1/2*(1+tanh(25*(\x-1/2)))};

    \draw [dotted, thick] (axis cs:1/2,0)-- (axis cs:1/2,1);
\end{axis}
\node[label=below:{$\epsilon^*$}] at (1.65,0.2) {};
\node[label=above:{$\xleftrightarrow{\!\!\Theta(1\!/\!{\sqrt{N}})\!\!}$}] at (1.5,2.8) {};

\end{tikzpicture}
    \end{center}
  \end{subfigure}%
\vspace{-.2cm}
    \caption{Left: Sharp transition of the block-MAP error. From \cite{tillich2000discrete} we know that
	 the transition width is $O(1/\sqrt{d_{\rm min}})$. Right: The {\em optimal} transition width is $\Theta(1/\sqrt{N})$.}
    \label{fig:transition}
\end{figure}

\noindent\textbf{Sharp Transition for the Block-MAP Error.}  Let us view $P_{\rm MAP}(\epsilon)$ as a function of the channel parameter $\epsilon$. Consider first transmission over the BEC. In this case, it is not hard to see that $P_{\rm MAP}(\epsilon)$ is an increasing function of $\epsilon$ for $\epsilon \in [0,1]$ with $P_{\rm MAP}(\epsilon=0) = 0, P_{\rm MAP}(\epsilon=1)=1$. Furthermore, the function $P_{\rm MAP}$ exhibits a sharp transition behaviour~\cite{tillich2000discrete}: Let $\epsilon^*$ be such that $P_{\rm MAP}(\epsilon^*) = \frac 12$. Then, around $\epsilon = \epsilon^*$, the value of $P_{\rm MAP}$ jumps from ``almost zero" to ``almost one" and the transition width is of oder $O(1/\sqrt{d_{\rm min}})$. We refer to Fig.~\ref{fig:transition} for a schematic illustration. The same picture holds true when the transmission channel is a BSC($\epsilon$). More precisely, we have the following theorem from \cite{tillich2000discrete}.
\begin{theorem}\label{TZ}
Let $P_{\rm MAP}$ be the block-MAP error for transmission of a linear code $\code$ with minimum distance $d_{\min}$ over BEC($\epsilon$). We have
\begin{align*}
& P_{\rm MAP}(\epsilon) \leq \Phi(\sqrt{2d_{\rm min}}(\sqrt{-\ln \epsilon^*} - \sqrt{-\ln \epsilon})) \,\,\text{ for } 0< \epsilon<\epsilon^*,\\
& P_{\rm MAP}(\epsilon) \geq \Phi(\sqrt{2d_{\rm min}}(\sqrt{-\ln \epsilon^*} - \sqrt{-\ln \epsilon})) \,\,\text{ for }  \epsilon^*<\epsilon<1,
\end{align*}
where $\epsilon^*$ is defined by $P_{\rm MAP}(\epsilon^*) = \frac{1}{2}$ and $\Phi$ stands for the Gaussian cumulative distribution, i.e., $\Phi(x) = \int_{-\infty}^x \frac{1}{\sqrt{2\pi}}e^{-u^2/2} du$.

Furthermore, when the transmission is over BSC($\epsilon$), we have for $\epsilon < \epsilon^*$:
\begin{align*}
& P_{\rm MAP}(\epsilon)\! \leq \!\Phi(\sqrt{d_{\rm min}}(\sqrt{-\ln (1-\epsilon^*)} \!-\! \!\!\sqrt{-\ln(1- \epsilon)})),
\end{align*}
and for $\epsilon^* < \epsilon < \frac 12$:
\begin{align*}
& P_{\rm MAP}(\epsilon)\! \geq \!\Phi(\sqrt{d_{\rm min}}(\sqrt{-\ln (1-\epsilon^*)}\! - \!\!\sqrt{-\ln (1-\epsilon)})).
\end{align*}

\end{theorem}

\noindent\textbf{Optimal Transition Width and its Implications.}
Theorem~\ref{TZ} implies that when the code has linear minimum
distance (e.g., random codes or most LDPC codes) then the transition width
is $O(1/\sqrt{N})$. For Reed-Muller (RM) codes, which
have minimum distance $O(\sqrt{N})$ if we consider elements of fixed rate and increasing length, the implied
transition width is $O(N^{-\frac 14})$.

This suggests the following question: What is the optimal scaling of the transition
width (i.e., how ``fast'' can the transition be) in terms of the
blocklength $N$?   It is not hard to see that the optimal transition
width is $\Theta(1/\sqrt{N})$, see Fig.~\ref{fig:transition}.  An intuitive argument for this
(e.g., for the BEC) is that, for any $\epsilon$, the number of
channel erasures is with high probability smeared out over the window
$[N\epsilon - \Theta(\sqrt{N}), N\epsilon + \Theta(\sqrt{N})]$. As
a result, one cannot expect a drastic change in $P_{\rm MAP}$ between
$\epsilon^*$ and $\epsilon^* + o(\frac{1}{\sqrt{N}})$. Let us
formally state and prove this result in the following proposition.

\begin{prop}
Let $P_{\rm MAP}$ be the block-MAP error for transmission of a
linear code $\code$ with minimum distance $d_{\min}$ over the
BEC($\epsilon$) (or the BSC($\epsilon$)). For an arbitrary $\delta
\in (0,1/2)$ let $\epsilon_1 (\epsilon_2$) be such that $P_{\rm
MAP}(\epsilon_1) = \delta$ ($P_{\rm MAP}(\epsilon_2) = 1-\delta$).
Then, there exists a constant $B(\delta) > 0$, independent of the
choice of the code, such that $\epsilon_2 - \epsilon_1 \geq
B(\delta)/\sqrt{N}$.
\end{prop}
\begin{proof}
The proof follows from the fact that the derivative of the product
measure for every monotone property (e.g. $d P_{\rm
MAP}(\epsilon)/d\epsilon$) is at most of order $O(\sqrt{N})$ (see
for example~\cite[Corollary 9.16]{boucheron2013concentration}).
\end{proof}

\section{Main Statement}
\begin{theorem} \label{BEC}
Let $\code$ be a binary linear code of length $N$, dimension $K$,
and with weight distribution $A(w)$. Consider transmission over the binary erasure channel with parameter
$0 < \epsilon^* < 1$, where $\epsilon^*$ is such that
\begin{align*}
P_{\rm MAP}(\epsilon^*)=P\{\hat{x}(Y) \neq X\} = \frac12.
\end{align*}
Then, for any $1\leq W \leq N$ the following holds. For $\epsilon < \epsilon^*$:
\begin{align*}
P_{\rm MAP}(\epsilon) \leq &\Phi \left(\sqrt{ W}(\sqrt{-\ln(\epsilon^*)}-\sqrt{-\ln(\epsilon)})\right)\!\\\ 
&+ 2\sqrt{-\log \epsilon} \, \sqrt{W} A(W, \epsilon^*),
\end{align*}
and for  $\epsilon^* < \epsilon < 1$:
\begin{align*}
P_{\rm MAP}(\epsilon) \geq &\Phi \left(\sqrt{ W}(\sqrt{-\ln(\epsilon^*)}-\sqrt{-\ln(\epsilon)})\right)\!\\\ 
&- 2\sqrt{-\log \epsilon^*} \sqrt{W} A(W, \epsilon).
\end{align*}
\end{theorem}
\begin{theorem}\label{BSC}
Let $\code$ be a binary linear code of length $N$, dimension $K$,
and with weight distribution $A(w)$. Consider transmission over the binary symmetric channel with parameter
$0 < \epsilon^* < 1$, where $\epsilon^*$ is such that
\begin{align*}
P_B(\epsilon^*)=P\{\hat{x}(Y) \neq X\} = \frac12.
\end{align*}
Then,  for any $1 \leq W \leq N$ following holds.
For $\epsilon < \epsilon^*$:
\begin{align*}
P_{\rm MAP}(\epsilon)  \leq \Phi&\left(\frac{\sqrt{W}}{2}(\sqrt{-\ln(1-\epsilon^*)}-\!\!\sqrt{-\ln(1-\epsilon)})\right)\\&  + 
4\sqrt{-\log \epsilon} \sqrt{W} A(W,\epsilon^*),
\end{align*}
and for $ \epsilon^* < \epsilon < \frac12$:
\begin{align*}
P_{\rm MAP}(\epsilon)  \geq  \Phi&\left(\frac{\sqrt{ W}}{2}(\sqrt{-\ln(1-\epsilon^*)}-\!\!\sqrt{-\ln(1-\epsilon)})\right) \\
&- 4\sqrt{-\log \epsilon^*} \sqrt{W} A(W,\epsilon).
\end{align*}
\end{theorem}
\noindent\textbf{Fast Transition for RM Codes.} One immediate
implication of Theorems~\ref{BEC} and \ref{BSC} is that the transition
width of a code $\code$ is at most $O(1/\sqrt{W})$ provided that
$\sqrt{W} A(W,z)$ is small. For RM codes, we use the following
result from \cite[Lemma 4]{kudekar2016comparing} to conclude that
the transition width is at most $\Theta(1/N^{\frac12-\kappa})$, for
any $\kappa>0$.
\begin{lemma}
For any $z \in [0,1)$ and any $\kappa>0$ the following holds for RM codes. Let $W = N^{1-\kappa}$, then
\begin{align*}
A(W,z) \leq e^{-W\beta(\kappa,z)},
\end{align*}
where $\beta(\kappa, z)$ is strictly positive for any $z \in (0,1)$ and $\kappa>0$.
\end{lemma}

\textit{Proof of Theorem~\ref{BEC}:} Consider transmission over the
BEC($\epsilon$) with a linear binary code $\mathcal{C}$ of blocklength
$N$. Since the code is linear and the channel is symmetric we can
assume without loss of generality that we transmit the all-zero
codeword. Given two sequences $x,y \in \{0,1\}^N$, we say that $y$
covers $x$ if the support of $x$ is included in the support of $y$,
i.e., $x_i \leq y_i$ for any $1\leq i \leq N$. Recall that we assume that the block length is $N$.
It is therefore natural to assign
to the $N$ channel actions a binary $N$-tuple, henceforth called
the \emph{erasure pattern}, which has value $1$ in its $i$-th
position if and only if the $i$-th channel erases its input and $0$ otherwise. In
this way, the set of all the erasure patterns $\{0,1\}^N$  is endowed
with the product measure as its corresponding probability measure.
We use $\mu_\epsilon(\cdot)$ to denote such a probability measure,
i.e., for an erasure pattern $\omega$ we have $\mu_\epsilon(\omega)
= \epsilon^{w_H(\omega)}(1-\epsilon)^{N-w_H(\omega)}$.   Furthermore,
assuming the all-zero transmission, an erasure pattern $\omega \in
\{0,1\}^N$ causes a block-MAP error if and only if there exists at
least one non-zero codeword $\cw \in \code$ which is covered by
$\omega$. We define $\Omega$ to be the set of erasure patterns which
cause a block-error, i.e.,
\begin{equation}
\Omega\! =\! \{\erasure \in \{0,1\}^N\!:\! \erasure \text{ covers  at least one non-zero codeword}\}.
\end{equation}
As a result, we have
\begin{equation}
P_{\rm MAP}(\epsilon) = \mu_\epsilon(\Omega).
\end{equation}
 Also, let us define the boundary of $\Omega$ to be
\begin{equation}
\partial\Omega = \{ \erasure \in \Omega:  \exists \erasure' \notin \Omega, d_H(\erasure, \erasure') =1\},
\end{equation}
where $d_H$ denotes the Hamming distance.

By definition, if $\erasure \in \partial \Omega$ then $\erasure$
covers at least one non-zero codeword.  We argue from \cite{tillich2000discrete}  that it covers in
fact {\em exactly one} non-zero codeword, call this codeword $\cw$.
This is true since if $\erasure$ covers two distinct non-zero codewords, call
them $\cw$ and $\cw'$, then by linearity of the code, it also covers
the codeword $\cw''=\cw+\cw'$. Now note that for every position
$i$, $1 \leq i \leq N$, at least one of $\cw_i$, $\cw_i'$, and
$\cw_i''$ must be $0$ (since by construction each of these values
is the XOR of the other two). Therefore, no matter what position
of $\erasure$ we set from erasure to non-erasure, at least one of
these three codewords will still be covered. In other words,
$\erasure$ does not have a neighbour at distance $1$ in $\Omega^c$,
i.e., $\erasure \not \in \partial \Omega$.

Given the product measure on the erasure patterns, the Margulis-Russo
formula expresses the derivate of $\mu_\epsilon(\Omega)$ in terms
of the measure of the boundary of $\Omega$:
\begin{equation}  \label{boundary-derivative0}
\frac{d \mu_\epsilon (\Omega)}{d \epsilon} = \frac{1}{\epsilon} \int_{\erasure \in \Omega}  h_{\Omega}(\erasure) d \mu_\epsilon(\erasure),
\end{equation}
where
\begin{align} \nonumber
&h_\Omega(\erasure) = 0, &\text{if }  \erasure\notin \Omega,\\ \label{equ:h}
&h_\Omega(\erasure) = \bigl| \{ \erasure' \notin \Omega: d_H(\erasure, \erasure')=1 \} \bigr|,  &\text{if }   \erasure \in \Omega.
\end{align}

Let us now see how the quantity $h_\Omega$ can be lower-bounded for boundary patterns $\omega \in \Omega$.
Let $\cw$ be the unique non-zero codeword that is covered by the
boundary point $\erasure$.  We write $\cw(\erasure)$. We know that
the weight of $\cw(\erasure)$ is at least $d_{min}$, and every
erasure pattern $\erasure'$ which is equal to $\erasure$ except at
one position $i$ where $\cw_i=1$ is an element of $\bar{\Omega}$
and $d_H(\erasure, \erasure')=1$. Hence, $\erasure$ has at least
$d_{\min}$ neighbours in $\bar{\Omega}$ as claimed, or in other words
$h_{\Omega}(\erasure) \geq d_{\min}$ \cite{tillich2000discrete}.
We will now strengthen this bound and show that
for most boundary points $\erasure$, $h_\Omega(\omega)$ is considerably
larger.

Let us define the set $\Gamma_W \subseteq \partial \Omega$,
\begin{equation} \label{gamma_W}
\Gamma_W = \{\erasure \in \partial \Omega:  w_H(\cw(\erasure)) \geq W\}.
\end{equation}
Note that
\begin{equation} \label{gamma-h}
\forall \erasure \in \Gamma_W: h_\Omega(\erasure) \geq W.
\end{equation}
We can then write
\begin{align} 
	\int_{\erasure \in \Omega}\!\! h_{\Omega}(\omega) d\mu_\epsilon(\omega) &\!\ge \!\sqrt{W}\!\! \int_{\erasure \in \Omega}\!\!\sqrt{h_{\Omega}} d\mu_\epsilon -\! \sqrt{W}\!\!
\int_{\omega \in \Gamma_W^c}\!\!\!\!\!\! \sqrt{h_{\Omega}} d\mu_\epsilon, \label{equ:int_h_lower}
\end{align}
where $\Gamma_W^c$ denotes the set complement of $\Gamma_W$.  By~\cite[Theorem 2.1]{tillich2000discrete} for \textit{monotone increasing sets $\Omega$} we have
\begin{equation} \label{equ:squ_h}
\int_{\omega \in \Omega} \sqrt{h_{\Omega}(\omega)} d\mu_\epsilon(\omega) \ge {1\over \sqrt{-2\log
\epsilon}}\gamma(\mu_\epsilon(\Omega)),
\end{equation}
where $\gamma(x) = \phi(\Phi^{-1}(x))$, where $\phi$ and $\Phi$ are the pdf and the CDF of standard normal distribution. 
Let us now bound the right-most term in \eqref{equ:int_h_lower}. We can write
$$\int_{\omega \in \Gamma_W^c}\!\!\!\!\!\! \sqrt{h_{\Omega}} d\mu_\epsilon = \sum_{w=1}^{W-1} \sqrt{w} 
\mu_\epsilon\left(\{\erasure \in \partial\Omega: w_H(\cw(\erasure))  = w\}\right). $$
Also, 
\begin{align}
 & \mu_\epsilon(\{\erasure \in \partial\Omega: w_H(\cw(\erasure)) = w\})  \nonumber \\
& = \mu_\epsilon(\{\erasure \in \partial\Omega: \exists \cw \in \code \ni (\erasure \succ \cw) \wedge(w_H(\cw) = w)\})  \nonumber \\
& \leq \mu_\epsilon(\{\erasure: \exists \cw \in \code \ni (\erasure \succ \cw) \wedge (w(\cw) =w )\})  \nonumber \\
& = \mu_\epsilon(\bigcup_{\cw \in \code:  w(\cw)  = w} \{\erasure: \erasure \succ \cw\})  \nonumber \\
& \leq  \sum_{\cw \in \code: w(\cw) = w} \mu_\epsilon( \{\erasure: \erasure \succ \cw\}). \label{gamma_bound}
\end{align}
When the channel is a BEC($\epsilon$) the last step of the above expression can be bounded by $ A(w) \epsilon^w$ and thus we obtain
\begin{equation} \label{equ:bata_BEC}
\int_{\omega \in \Gamma_W^c}\!\!\!\!\!\! \sqrt{h_{\Omega}} d\mu_\epsilon \leq \sum_{w=1}^W \sqrt{w} 
A(w) \epsilon^w. 
\end{equation}
Now, by using \eqref{equ:int_h_lower}, \eqref{equ:squ_h}, and \eqref{equ:bata_BEC} we obtain
\begin{align} \nonumber
\int_{\erasure \in \Omega}\!\! h_{\Omega}(\omega) & d\mu_\epsilon(\omega) \\
& \!\ge 
\sqrt{W} \left( {1\over \sqrt{-2\log
\epsilon}} \gamma(\mu_\epsilon(\Omega)) - \sum_{w=1}^W \sqrt{w} 
A(w) \epsilon^w\right) \nonumber \\
&\!\ge 
\sqrt{W} \left( {1\over \sqrt{-2\log
\epsilon}} \gamma(\mu_\epsilon(\Omega)) - \sqrt{W} A(W,\epsilon)\right). \label{equ:der_lower}
\end{align}
Combining \eqref{boundary-derivative0}, and \eqref{equ:der_lower}, we obtain  that for any $1 < W < N$
\begin{equation} \label{equ:der_bound}
\frac{d \mu_\epsilon (\Omega)}{d \epsilon} \geq \frac{\sqrt{W}}{\epsilon}\left( {1\over \sqrt{-2\log
\epsilon}} \gamma(\mu_\epsilon(\Omega)) - \sqrt{W} A(W,\epsilon)\right).
\end{equation}
Now, consider a channel parameter $\epsilon>\epsilon^*$. We have
\begin{align}
\mu_{\epsilon}(\Omega) - \mu_{\epsilon^*}(\Omega) &= \int_{\epsilon^*}^{\epsilon} \frac{d \mu_{\bar{\epsilon}} (\Omega)}{d \bar{\epsilon}} d \bar{\epsilon} \nonumber \\
& \geq\sqrt{W} \int_{\epsilon^*}^{\epsilon}\frac{1}{\bar{\epsilon}}\left( \frac{\gamma(\mu_{\bar{\epsilon}}(\Omega))}{\sqrt{-2\log
\bar{\epsilon}}} - \sqrt{W} A(W,\bar{\epsilon})\right) d \bar{\epsilon} 
\label{equ:int_bound}
\end{align}
Define $c(x) = \sqrt{-\log x}$. For computing the above integral, we consider two cases: (i) If
$\sqrt{W} A(W,\epsilon) \geq \frac12 \gamma(\mu_{\epsilon}(\Omega))/c(\epsilon^*)$, then by using the inequality $\gamma(x) \geq x(1-x)$ we obtain that  $\mu_{\epsilon}(\Omega) \geq 1 - 2c(\epsilon^*) \sqrt{W} A(W,\epsilon)$. Hence the result of the theorem holds for this case. (ii)  If
$\sqrt{W} A(W,\epsilon) < \frac12\gamma(\mu_{\epsilon}(\Omega))/ c(\epsilon^*)$, then as $A(W,\bar{\epsilon})$ is an increasing function in $\bar{\epsilon}$ and $\gamma(x)$ is concave and symmetric around $x=1/2$, then for any $\bar{\epsilon} \in [\epsilon^*, \epsilon]$ we have $\sqrt{W} A(W, \bar{\epsilon}) \leq 1/2 \gamma(\mu_{\bar{\epsilon}}(\Omega))/c(\bar\epsilon) $. As a result, the quantity inside the integral in \eqref{equ:int_bound} will be lower bounded by $\frac{1}{2\bar\epsilon}  \gamma(\mu_{\bar{\bar\epsilon}}(\Omega))/\sqrt{-2\log(\bar\epsilon)}$. Now, by integrating this new lower bound we obtain the result of the Theorem (for more details see \cite{tillich2000discrete}).

The result of the Theorem for $\epsilon < \epsilon^*$ follows similarly as above.

\textit{Proof of Theorem~\ref{BSC}:} Consider now transmission over
the BSC($\epsilon$) with a linear binary code $\mathcal{C}$ of
blocklength $N$.  Similar to the proof of Theorem~\ref{BEC}, we can
assume the all-zero transmission. Also, we can naturally map the
set of $N$ channel usages to an \emph{error pattern} $\omega \in
\{0,1\}^n$, where a $1$ at position $i$ means that the $i$-th
channel has flipped its input. In this way, the set of error patterns
is endowed with the product measure, i.e., i.e. for an error pattern
$\omega$ we have $\mu_\epsilon(\omega) =
\epsilon^{w_H(\omega)}(1-\epsilon)^{N-w_H(\omega)}$. . We let
$\Omega$ to be the set of error patterns which cause a block-error,
i.e.,
\begin{equation}
\Omega = \{ \erasure \in \{0,1\}^N: \exists \cw \in \mathcal{C}: \cw \neq 0, w_H(\omega + \cw) < w_H(\omega )  \}.
\end{equation}
In this regard, we have $P_{\rm MAP} = \mu_\epsilon(\Omega)$.
Also, let us define the boundary of $\Omega$ to be
\begin{equation}
\partial\Omega = \{ \erasure \in \Omega:  \exists \erasure' \notin \Omega, d_H(\erasure, \erasure') =1\}.
\end{equation}
The Margulis-Russo
formula \eqref{boundary-derivative0} expresses the derivate of $\mu_\epsilon(\Omega)$ in terms
of the function $h_{\Omega}$ (defined in \eqref{equ:h}) over the boundary of $\Omega$.
Now consider an error pattern $\omega$ in the boundary, i.e.,  $\omega \in \partial \Omega$. Then there exists at least one codeword, call it $x (\omega)$, for which $w_H(\omega + \cw) < w_H(\omega ) $. From  \cite{tillich2000discrete}, we  know that
\begin{equation}
h_\Omega(\omega) \geq \frac{w_{H}(x(\omega))}{2}.
\end{equation}
Hence, considering the set $\Gamma_W$ as in \eqref{gamma_W}, we have
\begin{equation}
\forall \omega \in \Gamma_W: h_\Omega(\omega) \geq \frac{W}{2}.
\end{equation}
We can now use the similar steps as for the derivation of \eqref{equ:der_bound} to show that for the case of the BSC we have
\begin{equation} \label{equ:der_bound_BSC}
\frac{d \mu_\epsilon (\Omega)}{d \epsilon} \geq \sqrt{W} \left( {\frac{1}{2 \sqrt{-2\log
\epsilon}}} \gamma(\mu_\epsilon(\Omega)) - 2\sqrt{W} A(W,\epsilon)\right).
\end{equation}
The rest of the proof now follows similarly to the case of the BEC.

\section{Estimating EXIT derivative via the Blowing-Up Lemma}

As above, we consider a linear code $\mathcal{C}$, $X\sim\mathrm{Uniform}(\mathcal{C})$ and denote by $Y^{\epsilon}$ be the result of passing $X$ through a BEC($\epsilon$), $0\leq \epsilon \leq 1$. For $\{i,j\}\in[n]$, define
\begin{align}
\mathcal{C}^{ij}_{00}&\triangleq\{c\in\mathcal{C} \ : \ (c_i,c_j)=(0,0)\},\nonumber\\
\mathcal{C}^{ij}_{01}&\triangleq\{c\in\mathcal{C} \ : \ (c_i,c_j)=(0,1)\},\nonumber\\
\mathcal{C}^{ij}_{10}&\triangleq\{c\in\mathcal{C} \ : \ (c_i,c_j)=(1,0)\},\nonumber\\
\mathcal{C}^{ij}_{11}&\triangleq\{c\in\mathcal{C} \ : \ (c_i,c_j)=(1,1)\},\nonumber.
\end{align}
For a vector $z\in\{0,1\}^n$ and a code $\mathcal{A}\subset\{0,1\}^n$ we define
\begin{align}
\mathcal{A}(z)\triangleq\{a\in \mathcal{A} \ : \ z\succ a\}.
\end{align}
We now define the following partition of $\{0,1\}^n$ w.r.t. the codebook $\mathcal{C}$ and the coordinates $\{i,j\}$:
\begin{align}
\mathcal{B}^{ij}_1&\triangleq\{z\in\{0,1\}^n \ : \ \mathcal{C}^{ij}_{01}(z)=\mathcal{C}^{ij}_{10}(z)=\mathcal{C}^{ij}_{11}(z)=\emptyset\},\nonumber\\
\mathcal{B}^{ij}_2&\triangleq\{z\in\{0,1\}^n \ : \ \mathcal{C}^{ij}_{01}(z)\neq\emptyset,\mathcal{C}^{ij}_{10}(z)=\mathcal{C}^{ij}_{11}(z)=\emptyset\},\nonumber\\
\mathcal{B}^{ij}_3&\triangleq\{z\in\{0,1\}^n \ : \ \mathcal{C}^{ij}_{10}(z)\neq\emptyset,\mathcal{C}^{ij}_{01}(z)=\mathcal{C}^{ij}_{11}(z)=\emptyset\},\nonumber\\
\mathcal{B}^{ij}_4&\triangleq\{z\in\{0,1\}^n \ : \ \mathcal{C}^{ij}_{11}(z)\neq\emptyset,\mathcal{C}^{ij}_{01}(z)=\mathcal{C}^{ij}_{10}(z)=\emptyset\},\nonumber\\
\mathcal{B}^{ij}_5&\triangleq\{z\in\{0,1\}^n \ : \ \mathcal{C}^{ij}_{01}(z)\neq\emptyset,\mathcal{C}^{ij}_{10}(z)\neq\emptyset,\mathcal{C}^{ij}_{11}(z)\neq\emptyset\}.\nonumber
\end{align}
Note that indeed $\cup_{k=1}^5 \mathcal{B}_k^{ij} =\{0,1\}^n$ due to the linearity of the code. To see this, note that if $z$ covers $c_1\in\mathcal{C}^{ij}_{01}$ and also $c_2\in\mathcal{C}^{ij}_{10}$, then it must also cover $c_3=c_1+c_2\in\mathcal{C}_{11}^{ij}$, since $\mathrm{supp}(c_1+c_2)\subset(\mathrm{supp}(c_1)\cup\mathrm{supp}(c_2))$. Using the same reasoning, and recalling that $\mathbf{0}\in\mathcal{C}_{00}^{ij}$ is covered by all $z\in\{0,1\}^n$, we see that each $z\in\{0,1\}^n$ can either 1)cover only codewords from $\mathcal{C}_{00}^{ij}$; 2)cover codewords from $\mathcal{C}_{00}^{ij}$ and one of the codebooks $\mathcal{C}_{01}^{ij}$, $\mathcal{C}_{10}^{ij}$, or $\mathcal{C}_{11}^{ij}$; 3)cover codewords from all four codebooks $\mathcal{C}_{00}^{ij}$, $\mathcal{C}_{01}^{ij}$,$\mathcal{C}_{10}^{ij}$, $\mathcal{C}_{11}^{ij}$.


Define the $n$-dimensional random vector $Z^{\epsilon}=Z^{\epsilon}(ij)$ such that $Z^\epsilon_i=Z^\epsilon_j=1$, and $Z^\epsilon_k\sim\mathrm{Bernoulli}(\epsilon)$ i.i.d, for $k\in[n]\setminus\{i,j\}$, and define the quantity
\begin{align}
\alpha_{ij}^\epsilon\triangleq\Pr\left(Z^{\epsilon}\in\mathcal{B}^{ij}_4|Z^{\epsilon}\in\mathcal{B}^{ij}_2\cup \mathcal{B}^{ij}_3\cup \mathcal{B}^{ij}_4 \right).
\end{align}

In the sequel, for a vector $x\in\mathcal{X}^n$ and a subset $A\subset[n]$, we denote $x_{\sim A}=x_{[n]\setminus A}$.
\begin{theorem}
Assume $I(X_i;Y^\epsilon_{\sim i})\in(\delta,1-\delta)$ for some $0<\delta< 1/2$. Then
\begin{align}
I(X_i;X_j|Y^\epsilon_{\sim i,j})\geq\frac{c(\delta)}{\sqrt{n}}\alpha^\epsilon_{ij},
\end{align}
where $c(\delta)$ is bounded away from $0$ if $\delta$ is bounded away from $0$.
\label{thm:twoinf}
\end{theorem}

Before proving Theorem~\ref{thm:twoinf}, let us demonstrate its implication. Define the EXIT function
\begin{align}
g(\epsilon)=-\frac{1}{n}\frac{d}{d\epsilon}I(X;Y^\epsilon)=\frac{1}{n}\sum_{i=1}^n H(X_i|Y^{\epsilon}_{\sim i}).
\label{eq:gexitDef}
\end{align}
We can further compute
\begin{align}
g'(\epsilon)&=\frac{d}{d\epsilon}g(\epsilon)\nonumber\\
&=\frac{1}{n}\sum_{i=1}^n \sum_{j\neq i} \frac{\partial}{\partial \epsilon_j} H(X_i|Y^{\epsilon_j}_j,Y^\epsilon_{\sim i,j})\bigg|_{\epsilon_j=\epsilon}\nonumber\\
&=-\frac{1}{n}\sum_{i=1}^n \sum_{j\neq i} \frac{\partial}{\partial \epsilon_j} I(X_i,Y^{\epsilon_j}_j|Y^\epsilon_{\sim i,j})\bigg|_{\epsilon_j=\epsilon}\nonumber\\
&=\frac{1}{n}\sum_{i=1}^n \sum_{j\neq i} I(X_i,X_j|Y^\epsilon_{\sim i,j}).\label{eq:gder}
\end{align}

\begin{theorem}
Let $\mathcal{C}$ be a $1$-transitive code. Then if $g(\epsilon)\in(\delta,1-\delta)$ for some $0<\delta
\leq 1/2$, then
\begin{align}
g'(\epsilon)\geq \frac{c(\delta)}{n^{3/2}}\sum_{i=1}^n \sum_{j\neq i}\alpha^\epsilon_{ij},
\end{align}
where $c(\delta)$ is bounded away from $0$ if $\delta$ is bounded away from $0$.
\label{thm:transitiveder}
\end{theorem}

\begin{proof}
For transitive codes $g(\epsilon)=1-I(X_i;Y_{\sim i})$ for all $i\in[n]$. Combining~\eqref{eq:gder} with Theorem~\ref{thm:twoinf}, gives the result.
\end{proof}

\begin{proof}[Proof of Theorem~\ref{thm:twoinf}]
For any $y\in\{0,1,?\}^{n-2}$ set
\begin{align}
\alpha_A(y)&\triangleq\Pr(X_i=0,X_j=0|Y_{\sim i,j}^{\epsilon}=y)\nonumber\\
\alpha_B(y)&\triangleq\Pr(X_i=0,X_j=1|Y_{\sim i,j}^{\epsilon}=y)\nonumber\\
\alpha_C(y)&\triangleq\Pr(X_i=1,X_j=0|Y_{\sim i,j}^{\epsilon}=y)\nonumber\\
\alpha_D(y)&\triangleq\Pr(X_i=1,X_j=1|Y_{\sim i,j}^{\epsilon}=y)\nonumber,
\end{align}
and $P(y)\triangleq[\alpha_A(y) \ \alpha_B(y) \ \alpha_C(y) \ \alpha_D(y)]$. Let $S=\{k_1,\ldots,k_{|S|}\}\subset[n]\setminus\{i,j\}$ be the locations of non-erased bits within $y$, and let $x^*$ be a codeword in $\mathcal{C}$ for which $x^*_S=y_S$ (such a codeword must always exist). Let $\bar{S}\triangleq[n]\setminus S$ be the erased bits, and note that in vector representation $\bar{S}$ is a random vector with distribution $Z^\epsilon$. Thus, given $y$, we have that the transmitted codeword $x$ is uniformly distributed on $x^*+\mathcal{C}(z)$, where $z$ is the vector representation of $\bar{S}$. Without loss of generality, we may assume that $x^*_i=x^*_j=0$. Thus, since $\mathcal{C}(z)$ is a subspace for any $z$, we have that
\begin{align}
P(y)=P(z)=\begin{cases}
P_1=
\left[1 \ 0 \ 0 \ 0\right] & z\in\mathcal{B}_1^{ij}\\
P_2=\left[\frac{1}{2} \ \frac{1}{2} \ 0 \ 0\right] & z\in\mathcal{B}_2^{ij}\\
P_3=\left[\frac{1}{2} \ 0 \ \frac{1}{2} \ 0\right] & z\in\mathcal{B}_3^{ij}\\
P_4=\left[\frac{1}{2} \ 0 \ 0 \ \frac{1}{2}\right] & z\in\mathcal{B}_4^{ij}\\
P_5=\left[\frac{1}{4} \ \frac{1}{4} \ \frac{1}{4} \ \frac{1}{4}\right] & z\in\mathcal{B}_5^{ij}
\end{cases}
\end{align}
Note that $I(X_i;X_j|Y^{\epsilon}_{\sim i,j}=y)=1$ if $P(y)=P(z)=P_4$ and $I(X_i;X_j|Y^{\epsilon}_{\sim i,j}=y)=0$ otherwise. Thus, defining $Q_k=\Pr(Z^\epsilon\in\mathcal{B}_k^{ij})$, $k\in[5]$, we have
\begin{align}
I(X_i;X_j|Y^{\epsilon}_{\sim i,j})=Q_4.
\end{align}
By inspection of the $5$ different possibilities for $P(y)$, we observe that
\begin{align}
&H(X_i|Y^\epsilon_{\sim i})=Q_1\cdot 0+Q_2\cdot 0+Q_3\cdot 1+Q_4\cdot \epsilon+Q_5\cdot 1.\nonumber
\end{align}
%
%
Consequently,
\begin{align}
I(X_i;Y^{\epsilon}_{\tilde i})&=1-(Q_3+\epsilon\cdot Q_4+Q_5)\nonumber\\
&=Q_1+Q_2+(1-\epsilon)\cdot Q_4,\nonumber
\end{align}
and by the theorem's assumption, we therefore have that
\begin{align}
Q_1+Q_2+(1-\epsilon)\cdot Q_4\in(\delta,1-\delta).\label{eq:probabilitiesconstraint}
\end{align}
We proceed by using the Blowing Up Lemma (see e.g. \cite[Theorem 5.3]{csiszar2011information}) 
to show that~\eqref{eq:probabilitiesconstraint} implies that $\eta\triangleq Q_2+Q_3+Q_4\geq\frac{c(\delta)}{\sqrt{n}}$ for some constant $c(\delta)$. 

Define the set $\Omega^{ij}=\mathcal{B}^{ij}_2\cup \mathcal{B}^{ij}_3\cup \mathcal{B}^{ij}_4\cup  \mathcal{B}^{ij}_5$ and let $\partial \Omega^{ij}$ be its boundary. Further, let $\mathcal{D}^{ij}=\mathcal{B}^{ij}_2\cup \mathcal{B}^{ij}_3\cup \mathcal{B}^{ij}_4$. The crucial observation is that $\partial \Omega^{ij}\subset D^{ij}$, as erasure of a single additional coordinate, which corresponds to changing the Hamming weight of the erasure pattern by $1$, can increase the conditional entropy of $(X_i,X_j)$ by at most one bit. Applying the blowing-up lemma (see e.g. \cite[Theorem 5.3]{csiszar2011information}), we therefore have that
\begin{align}
\eta=\Pr(Z^\epsilon\in \mathcal{D}^{ij})&\geq \Pr(Z^\epsilon\in \partial\Omega^{ij})\nonumber\\
&\geq\frac{a}{\sqrt{n}}\gamma\left(\Pr(Z^\epsilon\in \Omega^{ij}) \right),
\label{eq:blowup}
\end{align}
where $a=a(\epsilon)$ is a positive constant and $\gamma$ is as defined after \eqref{equ:squ_h}.
Invoking~\eqref{eq:probabilitiesconstraint}, we see that either $Q_2+Q_3+Q_4>\delta/2$ or $\Pr(Z^\epsilon\in\Omega^{ij})=1-Q_1\in(\delta/2,1-\delta/2)$. Thus, $\eta\geq c(\delta)/\sqrt{n}$ where
$c(\delta)=a\gamma(\delta/2)$. The theorem now follows since $Q_4=\alpha^{\epsilon}_{ij}\cdot \eta$.
\end{proof}

\bibliographystyle{IEEEtran}
\bibliography{lthpub,lth}

\end{document}